\documentclass[fleqn,usenatbib]{mnras}

\usepackage{newtxtext,newtxmath}

\usepackage[T1]{fontenc}

\DeclareRobustCommand{\VAN}[3]{#2}
\let\VANthebibliography\thebibliography
\def\thebibliography{\DeclareRobustCommand{\VAN}[3]{##3}\VANthebibliography}

\usepackage{graphicx}	
\usepackage{amsmath}	

\title[Impact of metallicity on rotation \& activity]{The impact of stellar metallicity on rotation and activity evolution in the Kepler field using gyro-kinematic ages}

\author[V. See et al.]{
Victor See$^{1}$\thanks{E-mail: victor.see@esa.int}\thanks{ESA research fellow},
Yuxi (Lucy) Lu$^{2,3}$,
Louis Amard$^{4}$,
Julia Roquette$^{5}$
\\
$^{1}$European Space Agency (ESA), European Space Research and Technology Centre (ESTEC), Keplerlaan 1, 2201 AZ Noordwijk, The Netherlands\\
$^{2}$Department of Astronomy, Columbia University, 550 West 120th Street, New York, NY, USA\\
$^{3}$American Museum of Natural History, Central Park West, Manhatten, NY, USA\\
$^{4}$D\'{e}partement d’Astrophysique/AIM, CEA/IRFU, CNRS/INSU, Univ.  Paris-Saclay \& Univ.  de Paris, 91191 Gif-sur-Yvette, France\\
$^{5}$D\'{e}partement d'Astronomie, Universit\'{e} de Gen\`{e}ve, Chemin Pegasi 51, 1290 Versoix, Switzerland
}

\date{Accepted XXX. Received YYY; in original form ZZZ}

\pubyear{2024}

\begin{document}
\label{firstpage}
\pagerange{\pageref{firstpage}--\pageref{lastpage}}
\maketitle

\begin{abstract}
In recent years, there has been a push to understand how chemical composition affects the magnetic activity levels of main sequence low-mass stars. Results indicate that more metal-rich stars are more magnetically active for a given stellar mass and rotation period. This metallicity dependence has implications for how the rotation periods and activity levels of low-mass stars evolve over their lifetimes. Numerical modelling suggests that at late ages more metal-rich stars should be rotating more slowly and be more magnetically active. In this work, we study the rotation and activity evolution of low-mass stars using a sample of Kepler field stars. We use the gyro-kinematic age dating technique to estimate ages for our sample and use the photometric activity index as our proxy for magnetic activity. We find clear evidence that, at late ages, more metal-rich stars have spun down to slower rotation in agreement with the theoretical modeling. However, further investigation is required to definitively determine whether the magnetic activity evolution occurs in a metallicity dependent way.
\end{abstract}




\section{Introduction}
\label{Sec:Intro}
The rotation and magnetic activity of low-mass stars ($M_\star \lesssim 1.3M_\odot$) evolve significantly over their lives. Observations of stars in open clusters of known ages show that their rotation slows by several orders of magnitude during their main sequence lifetimes \citep[e.g.][]{Meibom2009,Meibom2011,Barnes2016,Rebull2016,Douglas2016,Douglas2017,Douglas2019,Curtis2020,Dungee2022}. This is due to their magnetised stellar winds carrying angular momentum away from the star in a process known as magnetic braking. Since magnetic activity is correlated with rotation \citep[e.g][]{Wright2011}, magnetic activity undergoes a corresponding weakening as low-mass stars age. Numerous numerical evolution models exist in the literature that attempt to reproduce the observed rotation period and activity evolution \citep[e.g.][]{Gallet2013,Gallet2015,Brown2014,Matt2015,Tu2015,Johnstone2015Rotation,Johnstone2021,Amard2016,Amard2019,Amard2020RotEvo,Blackman2016,vanSaders2016,Gondoin2017,Ardestani2017,See2018,Garraffo2018,SpadaLanzafame2020,Breimann2021,Gossage2021}. In general, these models are able to reproduce the overall trends although none are able to reproduce every observed feature. 

A crucial component of these types of evolution models is the spin-down torque, i.e. specifying the rate at which stars lose angular momentum as a function of their properties. Parameter studies using magneto-hydrodynamic simulations have shown that the rate at which stars lose angular momentum through their winds depends on the magnetic field strength, magnetic field topology and mass-loss rate of the star \citep{Matt2012,Reville2015,Garraffo2016,Pantolmos2017,Finley2017,Finley2018}. Therefore, to model the evolution of low-mass stars, one needs an understanding of their magnetic and wind properties. 

Stellar magnetic fields and mass-loss from winds are both, ultimately, products of the stellar dynamo. Although the process of magnetic field generation by dynamos is still not precisely understood, it is clear that it involves the interaction of rotation and convective motions within stellar interiors \citep[e.g.][]{Brun2017}. The parameter that, therefore, seems to govern the generation of magnetic activity is the stellar Rossby number (defined here as the rotation period over the convective turnover time, ${\rm Ro}=P_{\rm rot}/\tau$). Indeed, measurements of the magnetic fields of low-mass stars indicate that they are well parameterised by the Rossby number, with stars with smaller Rossby numbers having stronger magnetic fields up to a saturation value \citep[e.g.][]{Reiners2009,Vidotto2014,See2015,2019ZBvsZDI,See2019Geom,Kochukhov2020,Reiners2022}. However, determining whether mass-loss rates have a similar scaling with Rossby number is much harder. Observationally constraining the mass-loss rates of low-mass stars is extremely difficult due to how rarefied their winds are. From a theoretical standpoint, one might expect that mass-loss rates should follow a similar dependence on the Rossby number as magnetic field strengths since stellar wind driving depends strongly on a star's magnetic properties \citep[e.g.][]{Suzuki2013,Shoda2023}. Additionally, chromospheric and coronal forms of magnetic activity also scale similarly with Rossby number \citep[e.g.][]{Noyes1984,Saar1999,Pizzolato2003,Mamajek2008,Stelzer2016,Newton2017,Wright2018,Boudreaux2022}. Since stellar winds also arise from coronal heating, this adds to the expectation that mass-loss rates should also scale in this manner with Rossby number.

In recent years, attention has turned to understanding how stellar metallicity affects the rotation and activity evolution of low-mass stars. The theoretical expectation is that, at fixed stellar mass and rotation period, more metal-rich stars should have deeper convection zones and longer convective turnover times \citep{vanSaders2013,Karoff2018,Amard2019,Amard2020RotEvo,Claytor2020}, resulting in smaller Rossby numbers and higher levels of activity. Investigations using the photometric variability amplitude \citep{Karoff2018,Reinhold2020,See2021} and the flaring luminosity \citep{See2023} as magnetic activity proxies indicate that more metal-rich stars are, indeed, more magnetically active. Although this link between metallicity and activity has only been verified for a limited number of activity proxies, it is reasonable to expect that all forms of magnetic activity should have a similar link with metallicity, including the products of the stellar dynamo that govern angular momentum loss. Theoretical modeling suggests that, if angular momentum loss is metallicity dependent, metal-rich stars should be rotating slower than metal-poor stars at late ages for a given age and stellar mass \citep{Amard2020RotEvo}. An analysis of the rotation periods of stars in the Kepler field by \citet{Amard2020Kepler} found that metal-rich stars do rotate slower than metal-poor stars on average. A priori, this result could indicate that older stars are more metal-rich than younger stars since older stars rotate more slowly. However, such a trend is not supported by galactic studies \citep[e.g.][]{Haywood2013}. As such, the result of \citet{Amard2020Kepler} strongly suggests that angular momentum-loss is indeed metallicity dependent.

Although the result of \citet{Amard2020Kepler} is strongly indicative of a link between angular momentum loss and metallicity, it is not completely conclusive. One parameter that these authors did not have access to in their study is the stellar age. Determining the ages of main sequence low-mass field stars, like those in the Kepler field, is extremely difficult. One of the most commonly used methods is isochrone fitting \citep[e.g.][]{Berger2020}. However, this method suffers from large uncertainties when used on low-mass field stars since their properties do not evolve significantly along the main sequence. Another method which can be used to determine the ages of main sequence field stars is gyrochronology \citep[e.g.][]{Barnes2003,Barnes2007,Garcia2014,Angus2015,Angus2019}. Gyrochronology makes use of the fact that the rotation period of a star monotonically increases as it ages along the main sequence. Therefore, one should be able to estimate a star's age from its rotation period. However, gyrochronology relations are typically empirically calibrated and, to date, none explicitly account for stellar metallicity in the calibrations. 

Recently, \citet{Lu2021} developed the gyro-kinematic age dating method (see also \citet{Angus2020}) to estimate the ages of $\sim$30,000 stars in the Kepler field. Gyro-kinematic age dating makes use of the result from the field of galactic kinematics that older populations of stars generally have larger velocity dispersions \citep[e.g.][]{Nordstrom2004,Holmberg2009,Yu2018}. This link is known as the age-velocity-dispersion relation (AVR). The central assumption of gyro-kinematic age dating is that groups of stars with similar properties, such as their effective temperatures, rotation periods \textbf{and metallicities}, have the same age. Therefore, one can estimate the age of a group of stars with similar properties from their velocity dispersion and an AVR. 

In this work, we study how metallicity affects the rotation and activity evolution of a sample of Kepler field stars using the gyro-kinematic age dating method. Our aim is to answer two main questions. Firstly, do more metal-rich stars really spin down more rapidly than metal-poor stars as predicted by \citet{Amard2020RotEvo}? Secondly, how does metallicity affect the magnetic activity evolution of low-mass stars? More metal-rich stars are more magnetically active at fixed mass and rotation but they are also predicted to spin-down more rapidly. Since slow rotation is associated with weaker magnetic activity, it is not clear whether metal-rich stars should be more or less magnetically active than metal-poor stars at late ages.

\begin{figure}
	\includegraphics[trim=0mm 10mm 0mm 0mm,width=\columnwidth]{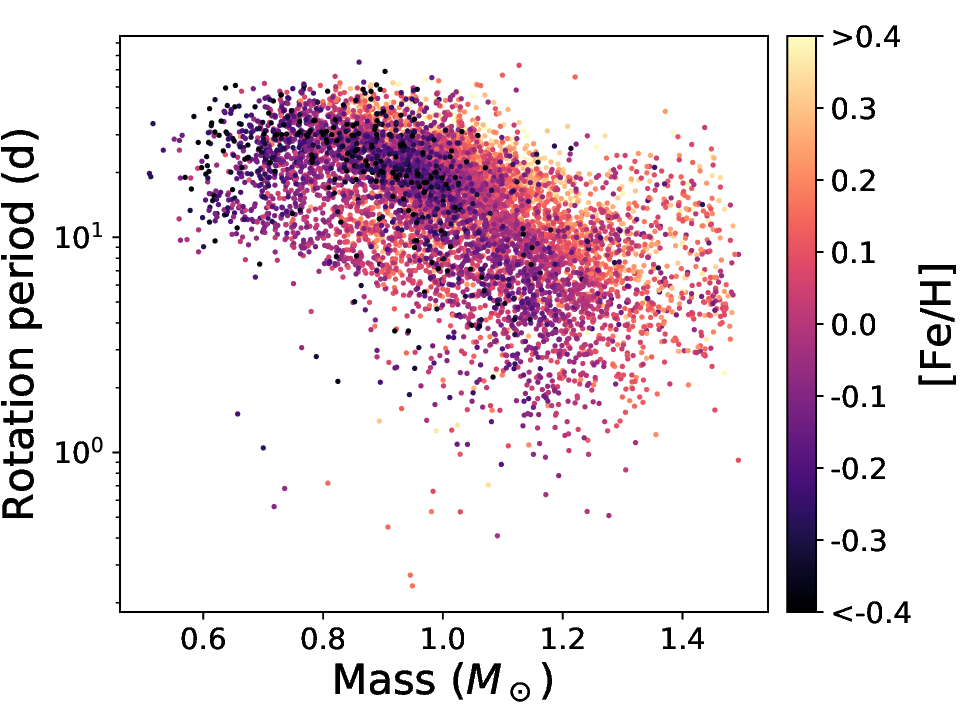}
    \caption{The rotation periods, masses and metallicities of our sample. The most metal-poor star in our sample is [Fe/H]=-0.89 and the most metal-rich is [Fe/H]=0.53. However, 97\% of our sample has metallicities between [Fe/H]=-0.4 and [Fe/H]=0.4 and we have truncated the colourbar at these values to increase the colour contrast of the plot.}
    \label{fig:Parameters}
\end{figure}

The rest of this paper is structured as follows. We describe how the stellar properties of our sample are determined in section \ref{Sec:Sample}, including a modification to the gyro-kinematic age dating method outlined by \citet{Lu2021}. In section \ref{sec:Results}, we present the rotation period, Rossby number and photometric activity index evolution as a function of age, paying special attention to the impact of metallicity. Finally, we present our conclusions in section \ref{sec:Conclusions}.

\section{Sample}
\label{Sec:Sample}
\subsection{Sample properties}
\label{Subsec:SamProperties}
In this work, we study a sample of well characterised main sequence field stars in the Kepler field. A number of different versions of this sample have been used in various studies \citep{Amard2020Kepler,See2021} with the most recent version being presented in \citet{See2023}. We briefly summarise the details of the sample here and refer the interested reader to \citet{See2023} for further details of how the sample was compiled. The main parameters of interest in this work are tabulated in table \ref{tab:SampleParams}.

\begin{figure}
	\includegraphics[trim=0mm 10mm 0mm 0mm,width=\columnwidth]{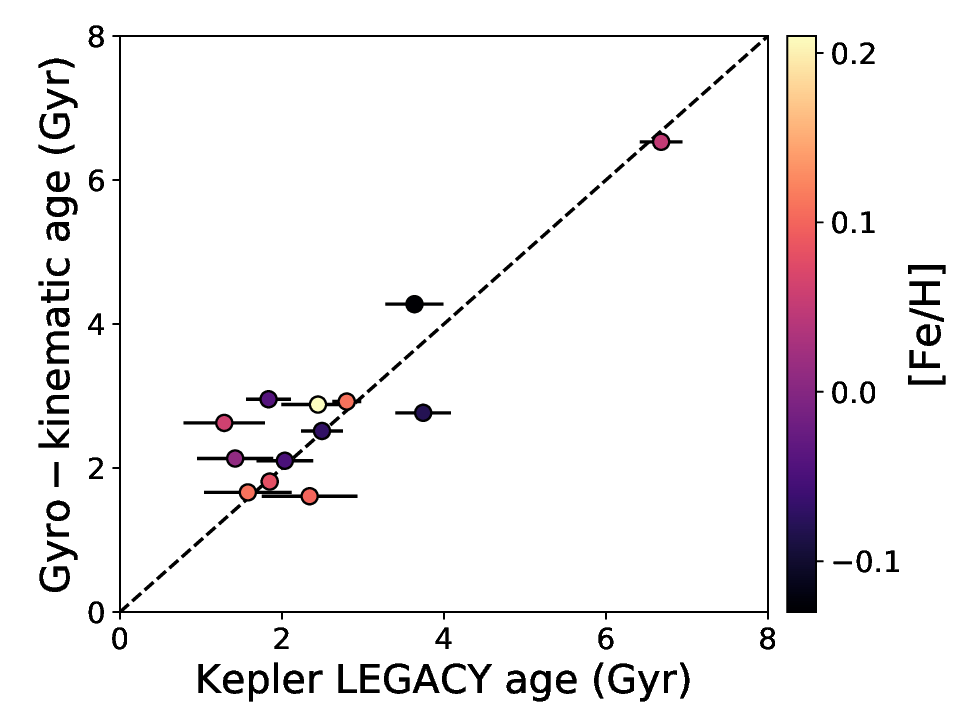}
    \caption{Estimated gyro-kinematic ages that result from our bin optimisation process against asteroseismic ages for a sample of LEGACY survey stars. The uncertainty on the LEGACY sample is estimated with the standard deviation of the ages resulting from different pipelines \citep{Silva2017}.
    The resulting $\chi^2$ is 3.13 and the optimised bin sizes for $T_{\mathrm{eff}}$, $\log_{10}(P_{\mathrm{rot}}$) and [Fe/H] are 83.67 K, 0.31 dex and 0.07 dex respectively.}
    \label{fig:OptimalBinsize}
\end{figure}

Rotation periods, $P_{\rm rot}$ (or equivalently, angular velocities, $\Omega = 2\pi / P_{\rm rot}$), for our sample are obtained from \citet{Santos2019}, \citet{Santos2021} and \citet{McQuillan2014} with preference given to periods from \citet{Santos2019} and \citet{Santos2021} over \citet{McQuillan2014} when periods for a star exist in multiple works. Metallicities, $\rm [Fe/H]$, are obtained from the APOGEE DR17 \citep{Abdurro'uf2022} and LAMOST DR7 surveys \citep{Luo2015,Liu2020,Du2021}. Masses, $M_\star$, and turnover times, $\tau$, are estimated using the grid of stellar strcuture models of \citet{Amard2019} and an adapted maximum-likelihood interpolation tool \citep{Valle2014}. Lastly, the photometric activity index, $S_{\rm ph}$, is adopted from \citet{Santos2019} and \citet{Santos2021}. Once these parameters were compiled, various cuts were applied to the Gaia DR3 colour-magnitude diagram to eliminate near equal mass binaries, eclipsing binaries, and evolved objects, as well as to ensure that the adopted parameters are of high quality. Full details of these cuts can be found in \citet{Amard2020Kepler} and \citet{See2023}. Our sample consists of 7752 stars whose main properties are shown in fig. \ref{fig:Parameters} (although the photometric activity index was only available for 7601 stars). 

\begin{table}
	\centering
	\caption{Parameters of interest in this work for our stellar sample. The full table, including references for each parameter, can be found online in a machine readable format.}
	\label{tab:SampleParams}
	\begin{tabular}{lcccccc}
		\hline
		$\rm KIC$ & $M_\star$ & $P_{\rm rot}$ & $\rm [Fe/H]$ & $\rm Age$ & $\tau$ & $S_{\rm ph}$\\
		& $(M_\odot)$ & (days) & (dex) & (Gyr) & (days) & (ppm) \\
		\hline
1026838 & 1.16 & 15.56 & 0.17 & 2.34 & 14.45 & 3.62e+02 \\
1027536 & 1.15 & 16.35 & 0.15 & 3.71 & 13.63 & 7.44e+01 \\
1161402 & 1.22 & 2.39 & -0.03 & 1.19 & 4.93 & 1.26e+03 \\
1161620 & 1.07 & 6.57 & 0.09 & 0.61 & 19.45 & 3.08e+03 \\
1162715 & 1.00 & 6.52 & 0.08 & 1.23 & 25.23 & 3.96e+03 \\
		\hline
	\end{tabular}
\end{table}

\begin{figure*}
	\includegraphics[trim=0mm 10mm 0mm 0mm,width=\textwidth]{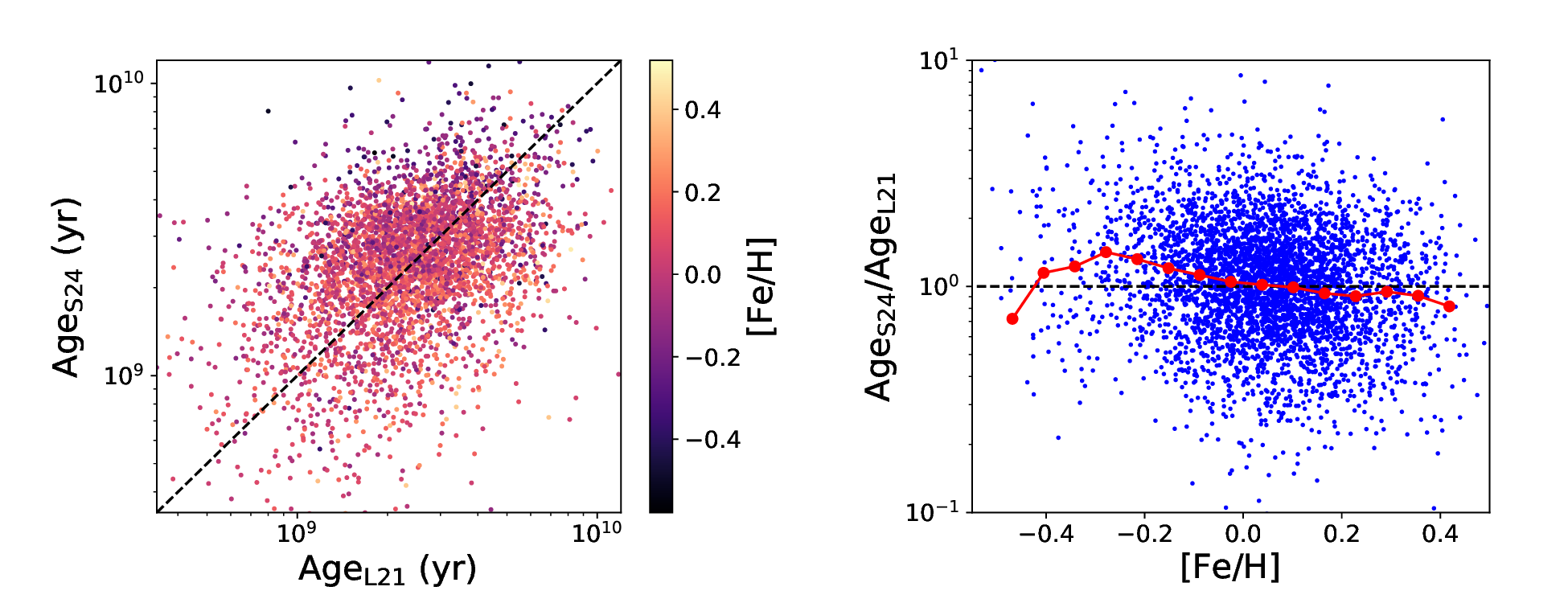}
    \caption{A comparison of the gyro-kinematic ages estimated in this work (subscripted S24) and those estimated in \citet[][subscripted L21]{Lu2021} for stars common to the samples of both works. Left: A direct comparison of the age estimates coloured by metallicity. Right: The ratio of the age estimates as a function of metallicity. The red points show a running average of the age ratio.}
    \label{fig:AgeComp}
\end{figure*}

\subsection{Gyro-kinematic ages}
\label{subsec:GyroAges}
The main update to this sample for this work is the addition of stellar ages calculated using the gyro-kinematic age dating method of \citet{Lu2021} which we briefly recap here. To calculate the gyro-kinematic age for a star, we first create a sub-sample of stars defined by a bin in effective temperature, rotation period and metallicity space centered on the star in question. In \citet{Lu2021}, the bin was created in absolute magnitude, effective temperature, rotation period and Rossby number space. However, for this work, we choose to include metallicity and drop absolute magnitude and Rossby number. Binning by metallicity is important as the goal of our study is to investigate how metallicity affects rotation and activity evolution. \citet{Lu2021} binned by absolute magnitude to account for evolved objects but we are focusing on main sequence stars in this work. Since we have already eliminated evolved objects from the sample as previously discussed, there is no need to bin by the absolute magnitude. Binning by Rossby number is also unnecessary as it should just be a function of the properties we have chosen to bin by, i.e. the effective temperature, rotation period and metallicity.

Once the sub-sample of stars has been determined, we calculate their vertical velocity dispersion, $\sigma_{v_z}$. This is given by 1.5 times the median absolute deviation of their vertical velocities, $v_z$. The vertical velocities of the stars are calculated from proper-motions and radial velocity measurements from Gaia DR3 \citep{Gaia2022}. We do this by transforming from the Solar system barycentric ICRS reference frame to Galactocentric Cartesian and cylindrical coordinates using {\tt astropy} \citep{astropy:2013, astropy:2018}. Once the vertical velocity dispersion has been calculated, it is then converted to an age estimate for the star in question using an AVR. \citet{Lu2021} used the AVR from \citet{Yu2018} that was determined using only metal-rich stars ([Fe/H]>-0.2). However, in this work, we use the AVR from \citet{Yu2018} that is determined using the full set of stars in their sample which encompasses both metal-rich and metal-poor stars. 

The size of the bins used to create the sub-samples is determined using an optimisation procedure. Gyro-kinematic ages are repeatedly calculated for our full sample of stars. Between each recalculation, the bin sizes in effective temperature, rotation period and metallicity space, are varied using a grid search. The adopted bin size for our final gyro-kinematic calculation is the one that produces the smallest $\chi^2$ when comparing our estimated gyro-kinematic ages with the asteroseismic ages of 13 stars from the Kepler LEGACY sample \citep{Silva2017}. Figure \ref{fig:OptimalBinsize} shows the final optimisation result, where the uncertainty on the LEGACY sample is estimated with the standard deviation of the ages resulting from different pipelines. The resulting $\chi^2$ is 3.13 and the optimised bin sizes for $T_{\mathrm{eff}}$, $\log_{10}(P_{\mathrm{rot}}$) and [Fe/H] are 83.67 K, 0.31 dex and 0.07 dex respectively.

Figure \ref{fig:AgeComp} shows a comparison of the gyro-kinematic age estimates from this work and the estimates from \citet{Lu2021} for stars common to the samples of both works. The left hand panel shows a direct comparison of the two estimates in log space. In general, they are correlated with a pearson correlation coefficient of 0.36 and a median absolute deviation of 0.9 Gyr. Perhaps surprisingly, there does not appear to be an obvious metallicity trend given that the age estimates of \citet{Lu2021} do not explicitly account for metallicity while ours do. The right hand panel of fig. \ref{fig:AgeComp} shows the ratio of the two age estimates as a function of metallicity. This is the same data as the left panel but presented in a slightly different format. We also plot a running average in red. This plot shows that there is a slight negative metallicity trend in stars with $\rm [Fe/H]\gtrsim -0.28$, where the majority of our sample is. For these stars, our gyro-kinematic age estimates are slightly larger than those of \citet{Lu2021} for metal-poor stars on average and slightly smaller than those of \citet{Lu2021} for metal-rich stars. For the full sample, there is a pearson correlation coefficient of -0.15 between the ratio of the age estimates in log space and the metallicity, indicating a weak correlation. This trend is small enough that it is not readily evident in the left hand plot and certainly much smaller than the overall scatter. This suggests that the dependence of gyro-kinematic ages on metallicity may be mostly captured by only binning in effective temperature, as they are not independent variables.

Lastly, we caution that the effects of weakened magnetic braking may add an additional level of uncertainty to the gyro-kinematic ages of some stars. Weakened magnetic braking is a phase of stellar rotation evolution, that occurs at late ages, where the standard spin-down laws appear to break down \citep[e.g.][]{vanSaders2016,Hall2021}. During this phase, angular momentum loss through stellar winds seems to lessen greatly or, possibly, even completely cease. Consequently, a star's rotation period may remain relatively unchanged over gigayear time-scales once it has entered the weakened magnetic braking phase. Since gyro-kinematic age dating assumes that stars with similar properties (including rotation period) have similar ages, gyro-kinematic ages derived for stars in the weakened magnetic braking phase are less reliable \citep{Lu2023}. In the main body of this work, we will analyse our full samples of stars including those in the weakened magnetic braking regime. However, we explore the effect stars in the weakened magnetic braking regime may have on our results in appendix \ref{app:WMB} and find that these stars do not affect our conclusions.

\begin{figure*}
	\includegraphics[trim=0mm 10mm 0mm 0mm,width=\textwidth]{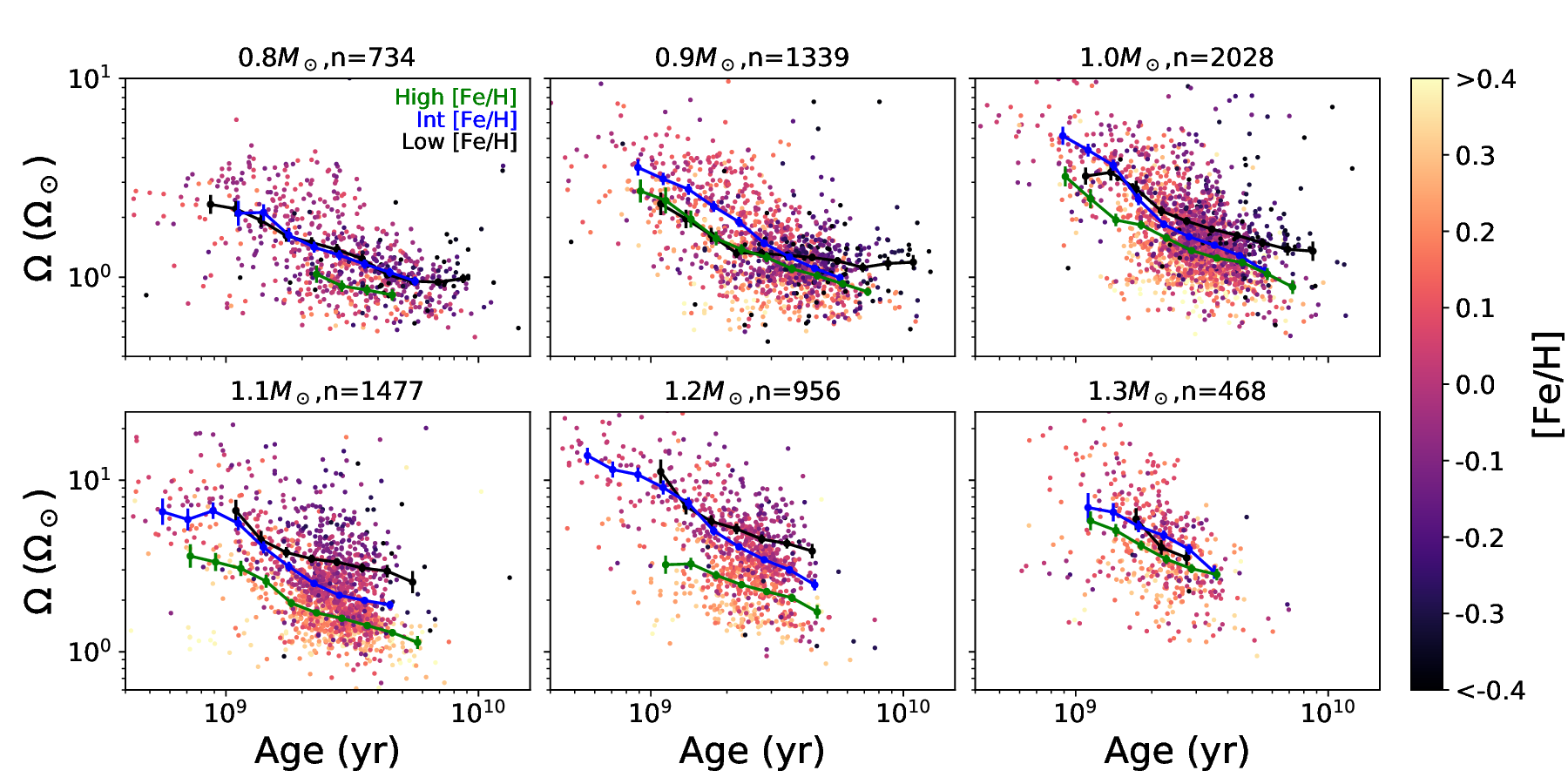}
    \caption{Angular velocity as a function of gyro-kinematic age and coloured by metallicity for 6 different mass bins. Each mass bin is 0.1 $M_\odot$ wide centered on the mass indicated above each panel. Running averages are also included for low (black points; $\rm [Fe/H]\leq -0.03$), intermediate (blue points; $\rm -0.03<[Fe/H]\leq +0.1$) and high (green points; $\rm [Fe/H]>+0.1$) metallicity stars. The standard error of each point in the running averages is indicated by a vertical bar. The number of stars within each mass bin is indicated above each panel. The most metal-poor star in our sample is [Fe/H]=-0.89 and the most metal-rich is [Fe/H]=0.53. However, 97\% of our sample has metallicities between [Fe/H]=-0.4 and [Fe/H]=0.4 and we have truncated the colourbar at these values to increase the colour contrast of the plot. Note that the y axis limits are different for the panels on the top and bottom rows.}
    \label{fig:PeriodAge}
\end{figure*}

\section{Results}
\label{sec:Results}
\subsection{Rotation evolution}
\label{subsec:RotResults}
\citet{Amard2020RotEvo} predicted that, at late ages, metal-rich stars should spin down to slower rotation than metal-poor stars for a given stellar mass using theoretical rotation evolution models. In this section, we test this prediction using the sample we described in section \ref{Sec:Sample}. Figure \ref{fig:PeriodAge} shows the angular velocity of our sample as a function of gyro-kinematic age and coloured by metallicity in six different mass bins. Each bin is 0.1$M_\odot$ wide and centered on the value indicated above each panel in fig. \ref{fig:PeriodAge}. Within each mass bin, the overall trend is that angular velocities are decreasing with age as expected. On average, stars are more rapidly rotating in higher mass bins which is also expected at late ages \citep[see e.g.][]{Matt2015}. Turning to the impact of metallicity, we generally see that more metal-rich stars have slower rotation than metal-poor stars at a given age. This qualitatively matches the prediction of \citet{Amard2020RotEvo}. 

To examine the trends more closely, we plot running averages in each panel for low ($\rm [Fe/H]\leq-0.03$; black points), intermediate ($\rm -0.03< [Fe/H]\leq +0.1$; blue points) and high metallicity ($\rm [Fe/H]> +0.1$; green points) stars. These running averages were calculated in log space and a data point is only plotted if more than 20 stars are available to calculate an average within a given age and metallicity bin. The boundaries of these metallicity bins were chosen to divide our full sample into thirds. We tested different values for the boundaries between the metallicity bins, e.g. ones that split the range of [Fe/H] values into thirds rather than ones that split the number of stars in each bin into thirds. However, we found that different boundary values do not appreciably change the trends in the running averages and, therefore, do not affect our conclusions. The size of the error bars are calculated as the standard deviation of the stellar angular velocities within each age and metallicity bin divided by the square root of the number of stars in that bin. The low and high metallicity running averages have been plotted with a small age offset of -0.01 dex and +0.01 dex respectively so that the error bars of the different running averages do not overlap.

A number of trends are evident in these running averages. The 1$M_\odot$, 1.1$M_\odot$ and 1.2$M_\odot$ mass bins all clearly show that, on average, metal-rich stars are spinning more slowly than metal-poor stars at a given age. This trend is not as clear in the 0.8$M_\odot$, 0.9$M_\odot$ and 1.3$M_\odot$ mass bins. For the 0.8$M_\odot$ bin, the trend may still be present since the running average corresponding to the highest metallicity stars has the slowest rotation, as expected. However, the intermediate and low metallicity running averages nearly lie exactly on top of each other. The 0.9$M_\odot$ mass bin is the most discrepant bin with the running average for the low metallicity stars crossing the running averages for the intermediate and high metallicity stars. The 1.3$M_\odot$ bin is similar to the 0.8$M_\odot$ bin in that the high metallicity running average has slower rotation than the other two running averages with the intermediate and low metallicity running averages overlapping somewhat. 

Several different factors likely contribute to the lack of clear metallicity trends in the 0.8$M_\odot$, 0.9$M_\odot$ and 1.3$M_\odot$ bins. The first factor concerns the 0.8$M_\odot$ and 0.9$M_\odot$ bins. Rotation evolution is predicted to be more sensitive to metallicity in higher mass stars \citep[see][]{Amard2020RotEvo}. Higher mass stars have thinner convection zones. Therefore, a given change in metallicity will result in a larger change in the convection zone depth and other convection zone properties for higher mass stars, in a fractional sense. This can be seen in figure 1 of \citet{Amard2020RotEvo} which shows that the convective turnover time is more sensitive to changes in metallicity in higher mass stars than in lower mass stars. This results in the dynamos of higher mass stars, as well its products, e.g. angular momentum loss through stellar winds, being more sensitive to metallicity than the dynamos of lower mass stars. The lack of a clear metallicity trend in the 0.8$M_\odot$ and 0.9$M_\odot$  bins may simply be a manifestation of this effect, whereby the predicted weak metallicity trend is being hidden by, e.g. measurement errors. A similar effect was hinted at in \citet{See2021} where we suggested that the photometric variability is more sensitive to changes in metallicity in high-mass stars than in low-mass stars. The second factor that may be hiding a metallicity trend in the running averages in the 0.8$M_\odot$ and 0.9 $M_\odot$ bins is that these bins probably suffer from an incompleteness problem since they lie at the faint end of our sample. They have fewer stars in them compared to some of the higher mass bins even though lower mass stars are more numerous than higher mass stars in the Milky Way. If these bins were more complete, the expected metallicity trend of slower rotation for more metal-rich stars might be more evident. Similarly, the lack of a clear metallicity trend in the 1.3$M_\odot$ bin may be due to the low number of stars in this bin. However, unlike the 0.8$M_\odot$ and 0.9$M_\odot$ bins, the small number of stars in this bin is likely to be intrinsic, rather than an incompleteness problem since higher mass stars are known to be less common than lower mass stars.

\begin{figure*}
	\includegraphics[trim=0mm 10mm 0mm 0mm,width=\textwidth]{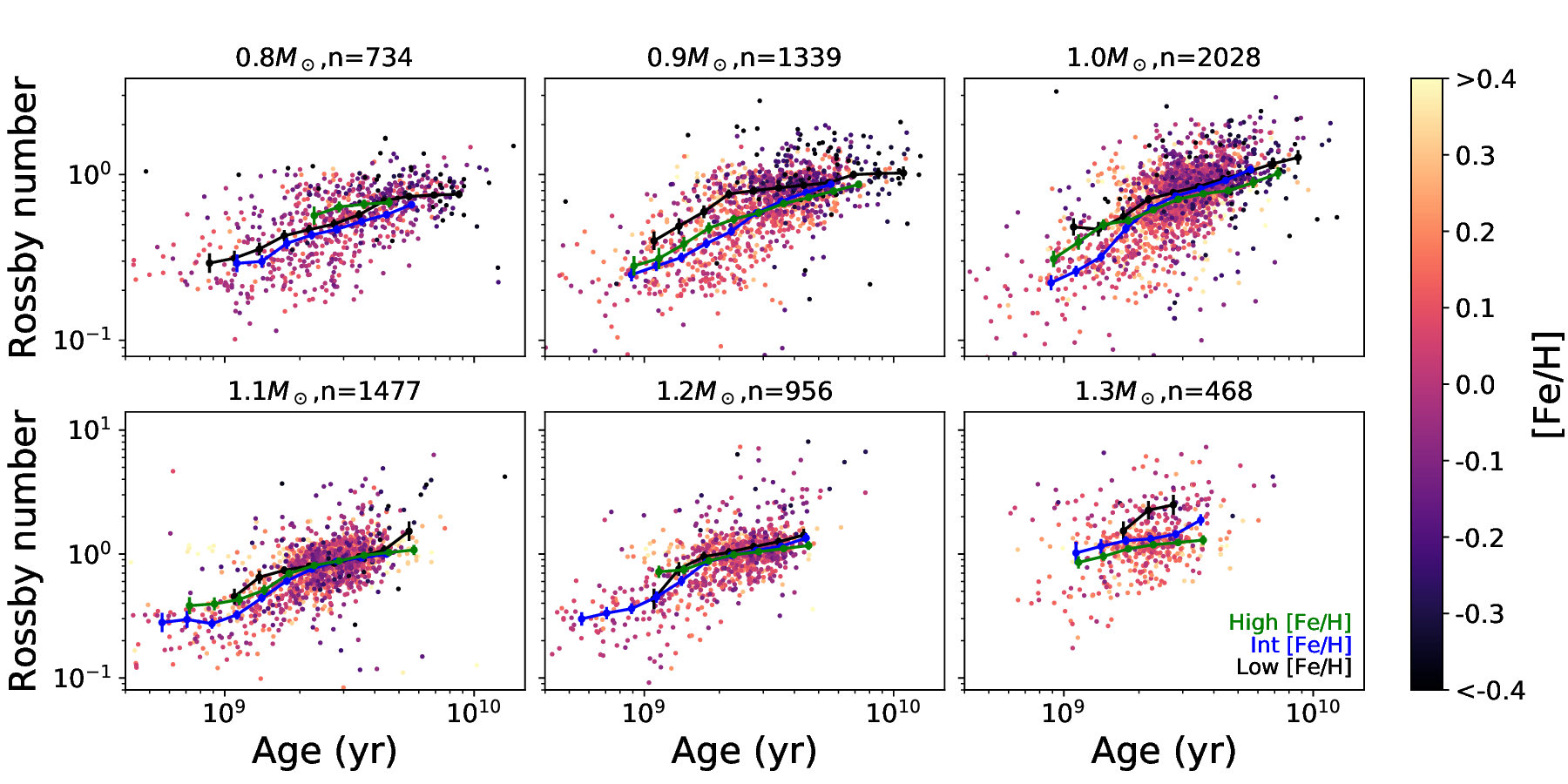}
    \caption{Same as fig. \ref{fig:PeriodAge} but for Rossby number instead of angular velocity.}
    \label{fig:RossbyAge}
\end{figure*}

\subsection{Rossby number evolution}
\label{subsec:RossResults}
In section \ref{subsec:ActResults}, we will look at the magnetic activity evolution of our sample. However, before we do so, we first look at the evolution of the Rossby number since this parameter is thought to determine how magnetically active a star is. Figure \ref{fig:RossbyAge} shows the same plot as fig. \ref{fig:PeriodAge} but for Rossby number rather than angular velocity. In each mass bin, the Rossby number increases as a function of age, with higher mass stars having larger Rossby numbers on average, as expected. 

In section \ref{subsec:RotResults}, we showed that more metal-rich stars generally have smaller angular velocities (or, equivalently, longer rotation periods) at fixed age and mass. One might, therefore, expect that the more metal-rich stars should have larger Rossby numbers since it is proportional to the rotation period. However, this is offset by the fact that the convective turnover time also increases with increasing metallicity. Given that the Rossby number is defined as the rotation period over the turnover time, it is not obvious how the Rossby number should scale with metallicity at fixed age and mass. A priori, it is not clear which of these two effects is larger or if they cancel out. The modelling of \citet{Amard2020RotEvo} suggests that the change in turnover time as a function of metallicity is larger than the change in rotation period at late ages and that the Rossby number should decrease with increasing metallicity (see their figure 5). However, the change in Rossby number that \citet{Amard2020RotEvo} predict is not particularly large over the range of metallicities that the majority of our sample is concentrated in ($\rm -0.2 \lesssim [Fe/H] \lesssim +0.3$). The running averages in fig. \ref{fig:RossbyAge} show some possible small hints that the Rossby number does decrease with increasing metallicity. For example, in the 1.3$M_\odot$ mass bin, the running average for the low metallicity stars is larger than the intermediate metallicity running average which is itself larger than the high metallicity running average. Additionally, the low metallicity running averages in the 0.9$M_\odot$ and 1$M_\odot$ bins are nearly always larger than the intermediate and high metallicity running averages. However, this is weak evidence, at best, of a metallicity trend. Additionally, there is no discernible metallicity trend in the 0.8$M_\odot$, 1.1$M_\odot$ and 1.2$M_\odot$ mass bins. Therefore, with the currently available data, we cannot confidently confirm or rule out whether the Rossby number evolution is metallicity dependent. It is worth bearing in mind that the gyro-kinematic age dating technique estimates ages in a statistical way and that the estimated ages for any individual star may be far from the true age. This may obscure any metallicity trends that exist, especially if the trends are relatively subtle as predicted by \citet{Amard2020RotEvo}. 

\begin{figure*}
	\includegraphics[trim=0mm 10mm 0mm 0mm,width=\textwidth]{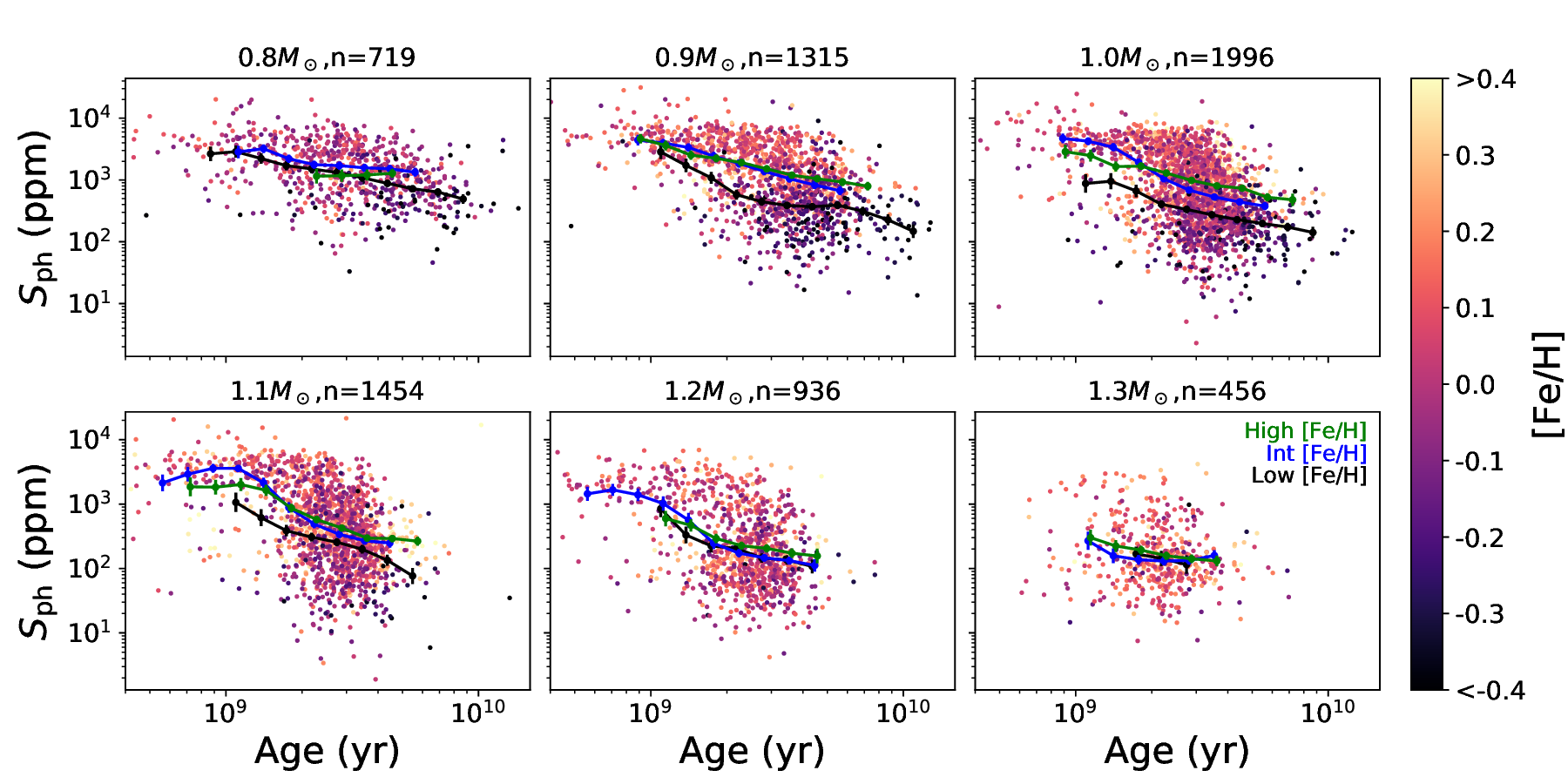}
    \caption{Same as fig. \ref{fig:PeriodAge} but for photometric activity index instead of angular velocity.}
    \label{fig:ActAge}
\end{figure*}

\subsection{Activity evolution}
\label{subsec:ActResults}
In this section, we look at how metallicity affects the magnetic activity evolution in low-mass stars. To do this, we use the photometric activity index values, $S_{\rm ph}$, calculated by \citet{Santos2019} and \citet{Santos2021} as our proxy for magnetic activity. This quantity is a measure of the variability in the light curve of a star on the rotational time-scale and has been shown to be a reasonable proxy of a star's magnetic activity \citep{Mathur2014}. It is determined by splitting a star's light curve into segments of length 5$P_{\rm rot}$ and calculating the standard deviation of each of these segments. The final photometric activity index is given by the average of these standard deviations after a correction for photon noise is applied.

Figure \ref{fig:ActAge} shows the same plot as fig. \ref{fig:PeriodAge} but for the photometric activity index rather than angular velocity. In each of the mass bins, the photometric activity index is decreasing with age which is the expected behaviour \citep[e.g.][]{Mathur2023}. Looking at the 0.9$M_\odot$, 1$M_\odot$ and 1.1$M_\odot$ bins, there are hints that, at a given age, more metal-poor stars have smaller photometric activity index values, although this trend is far from clear-cut. The running averages for the low metallicity stars are smaller than the running averages for the intermediate and high metallicity stars in these three mass bins. This is suggestive of a metallicity trend whereby more metal-rich stars have larger photometric activity index values for a given mass and age. If we assume that other forms or proxies of magnetic activity, such as mass-loss rates and magnetic field strengths, also follow the same metallicity trend as the photometric activity index, then the direction of this metallicity trend is consistent with our finding in section \ref{subsec:RotResults} that more metal-rich stars are spinning more slowly for a given mass and age. However, there is very little difference between the intermediate and high metallicity running averages in these bins making this interpretation less certain. In the remaining mass bins, i.e. the 0.8$M_\odot$, 1.2$M_\odot$ and 1.3$M_\odot$ bins, there is no obvious metallicity trend. Similarly to the Rossby number evolution in section \ref{fig:RossbyAge}, it is unclear if there is a metallicity dependence to the activity evolution shown in fig. \ref{fig:ActAge}.

A number of factors hamper our ability to reliably determine whether the photometric activity index evolution in fig. \ref{fig:ActAge} is metallicity dependent. Firstly, the photometric activity index is a relatively indirect tracer of magnetic activity. For example, a difference in stellar inclination can cause two stars with the same level of magnetic activity to have different levels of brightness variations in their light curves \citep{Mazeh2015,Shapiro2016} and, therefore, their photometric activity index values. This could add significant scatter to plots involving the photometric activity index that obscure the trends with metallicity that we are interested in. Indeed, we previously found that the scatter in the photometric variability amplitude, $R_{\rm per}$, which is a quantity similar to the photometric activity index, can be significant, even for a given Rossby number \citep{See2021}. Another complicating factor is that changes in metallicity can result in changes to the contrast of magnetic features \citep{Witzke2018,Witzke2020}. This would affect the amplitude of variability in a star's light curve and, therefore, its photometric activity index, in a way that is unrelated to changes in the stellar dynamo.

\section{Conclusions}
\label{sec:Conclusions}
In this work, we investigate how metallicity affects the evolution of the rotation, Rossby number and photometric activity index (which is our proxy for magnetic activity) of low-mass stars. We do this using a sample of nearly 8000 low-mass Kepler field stars. For this sample, we assemble literature values for the rotation period \citep{McQuillan2014,Santos2019,Santos2021}, metallicity \citep{Luo2015,Liu2020,Du2021,Abdurro'uf2022} and photometric activity index \citep{Santos2019,Santos2021}. We also calculate masses and convective turnover times using the stellar structure models of \citet{Amard2019} and ages using the gyro-kinematic age dating method of \citet{Lu2021}. We find a clear metallicity dependence to the rotation evolution. For a fixed mass and at a given age, more metal-rich stars are generally rotating more slowly than metal-poor stars in our sample. This is because, all else being equal, more metal-rich stars should have deeper convective envelopes, longer turnover times, smaller Rossby numbers and be more magnetically active \citep{See2021,See2023}. This results in stronger angular momentum loss through their magnetised winds. Additionally, the metallicity dependence of the rotation evolution appears to be stronger in higher mass stars. These results are in qualitative agreement with the modelling of \citet{Amard2020RotEvo} and provide the strongest evidence yet that angular momentum loss is a metallicity dependent process. 

While the interpretation for the metallicity dependence of the rotation evolution is relatively clear, the picture for the Rossby number and photometric activity index is much less so. In order for more metal-rich stars to spin down to slower rotation, one would expect that, at a given mass and age, more metal-rich stars should have smaller Rossby numbers and larger magnetic activities. Although our analysis shows some small hints that this is the case, it is far from conclusive. Unfortunately, with the available data, we cannot rule out the alternative scenario that there is no metallicity dependence to the Rossby number and photometric activity index evolution or even that they scale in the opposite direction as a function of metallicity. In the future, the question of how the Rossby number and magnetic activity evolution scales with metallicity could be better answered with larger sample sizes and by studying different activity indicators. In particular, studying different activity indicators would reduce some of the scatter present due to the fact that the photometric activity index is a comparatively indirect tracer of magnetic activity. However, it is not clear when such studies will be possible. In comparison to the photometric activity index, other magnetic activity indicators are comparatively difficult to measure. As such, it is challenging to assemble a large enough sample to conduct this type of study with other activity indicators.

\section*{Acknowledgements}
We thank the anonymous referee for comments that improved this work. V.S. acknowledges support from the European Space Agency (ESA) as an ESA Research Fellow. Y.L. acknowledges support from ESA through the Science Faculty of the European Space Research and Technology Centre (ESTEC). L.A. acknowledges support from the Centre National des Etudes Spatiales (CNES) through the PLATO/AIM grant. J.R. acknowledges funding from the European Union’s Horizon 2020 research and innovation program (grant agreement No.101004141, NEMESIS).

\emph{Software:} \texttt{matplotlib} \citep{Hunter2007}, \texttt{numpy} \citep{Harris2020}, \texttt{scipy} 
\citep{Virtanen2020}, \texttt{pandas} \citep{mckinney-proc-scipy-2010,pandas2023} \texttt{TOPCAT} \citep{Taylor2005}

\section*{Data Availability}
The data used throughout this work, i.e. the data contained in table \ref{tab:SampleParams}, will be made available via VizieR upon publication.

\appendix
\section{Impact of excluding stars in the weakened magnetic braking regime}
\label{app:WMB}
In this appendix, we explore whether stars in the weakened magnetic braking regime affect our results and conclusions. As discussed in section \ref{subsec:GyroAges}, gyro-kinematic ages for stars in the weakened magnetic braking regime are likely to have an additional level of uncertainty. Although it is still unclear what causes weakened magnetic braking and when exactly it occurs, recent modelling suggests that stars enter the weakened magnetic breaking phase once they reach a critical Rossby number of ${\rm Ro_{crit} = 0.91 Ro_\odot}$ \citep{Saunders2023}. Using a solar Rossby number of $\rm Ro_\odot = 1.26$, the value of which has been derived in the same way as for the stars in the rest of our sample, this corresponds to a critical Rossby number of ${\rm Ro_{crit} = 1.14}$. We find that 18\% of stars in our sample have Rossby numbers larger than this $\rm Ro_{crit}$.

\begin{figure*}
	\includegraphics[trim=0mm 10mm 0mm 0mm,width=\textwidth]{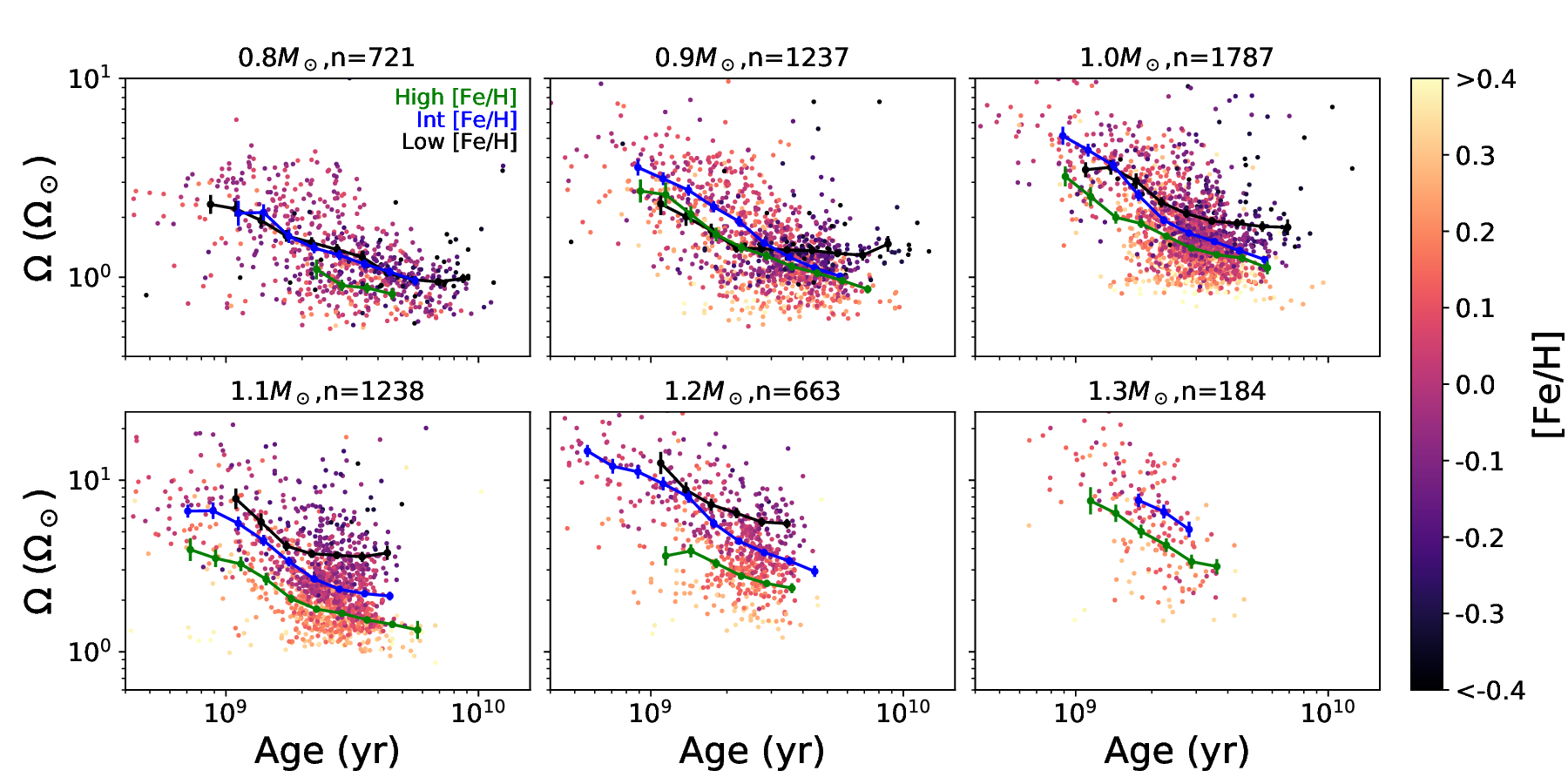}
    \caption{Same as fig. \ref{fig:PeriodAge} but only for stars that have not yet reached the weakened magnetic braking regime, i.e. those with $\rm Ro \le 1.14$.}
    \label{fig:PeriodAge_RossCut}
\end{figure*}

\begin{figure*}
	\includegraphics[trim=0mm 10mm 0mm 0mm,width=\textwidth]{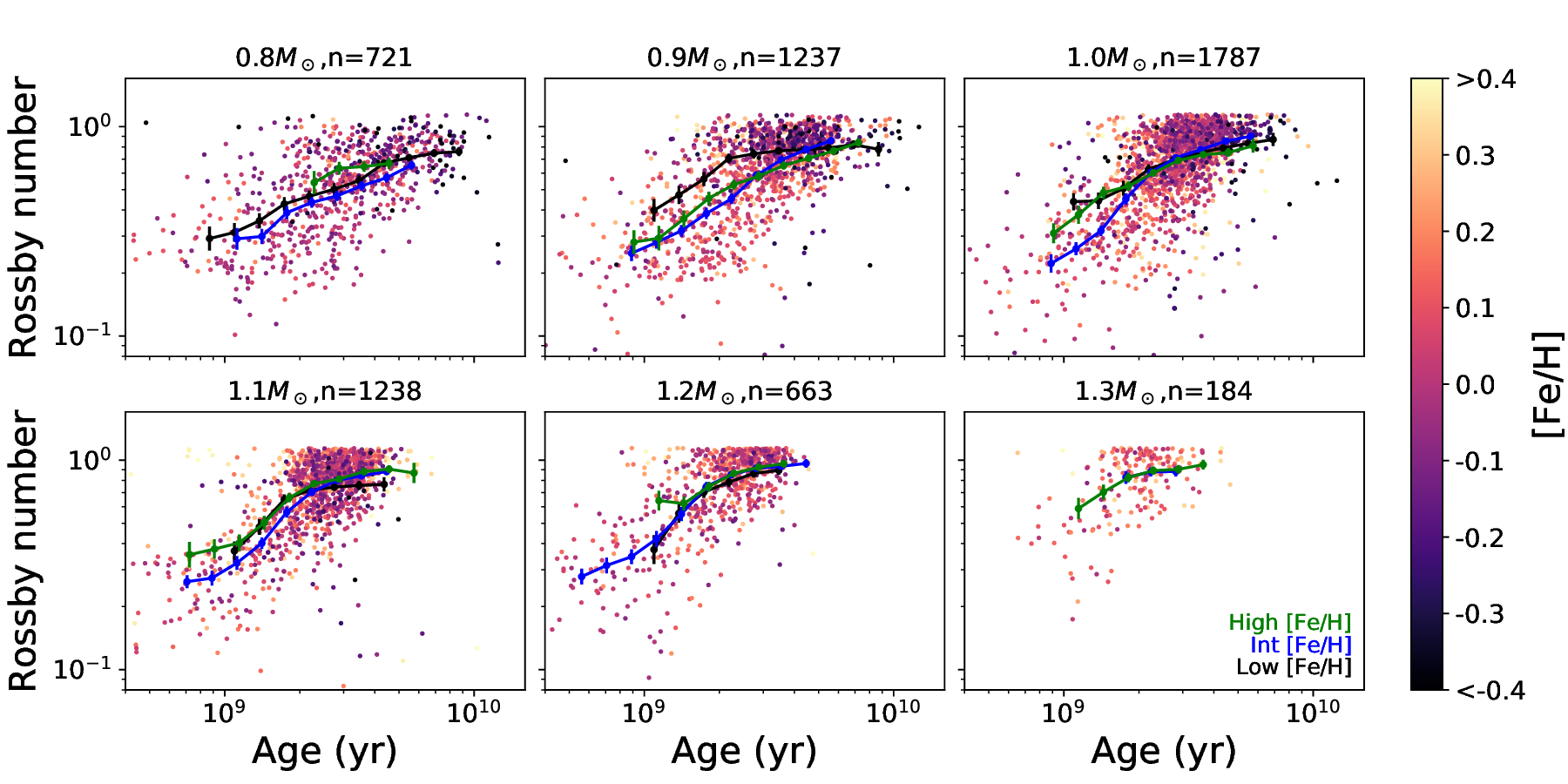}
    \caption{Same as fig. \ref{fig:RossbyAge} but only for stars that have not yet reached the weakened magnetic braking regime, i.e. those with $\rm Ro \le 1.14$.}
    \label{fig:RossbyAge_RossCut}
\end{figure*}

\begin{figure*}
	\includegraphics[trim=0mm 10mm 0mm 0mm,width=\textwidth]{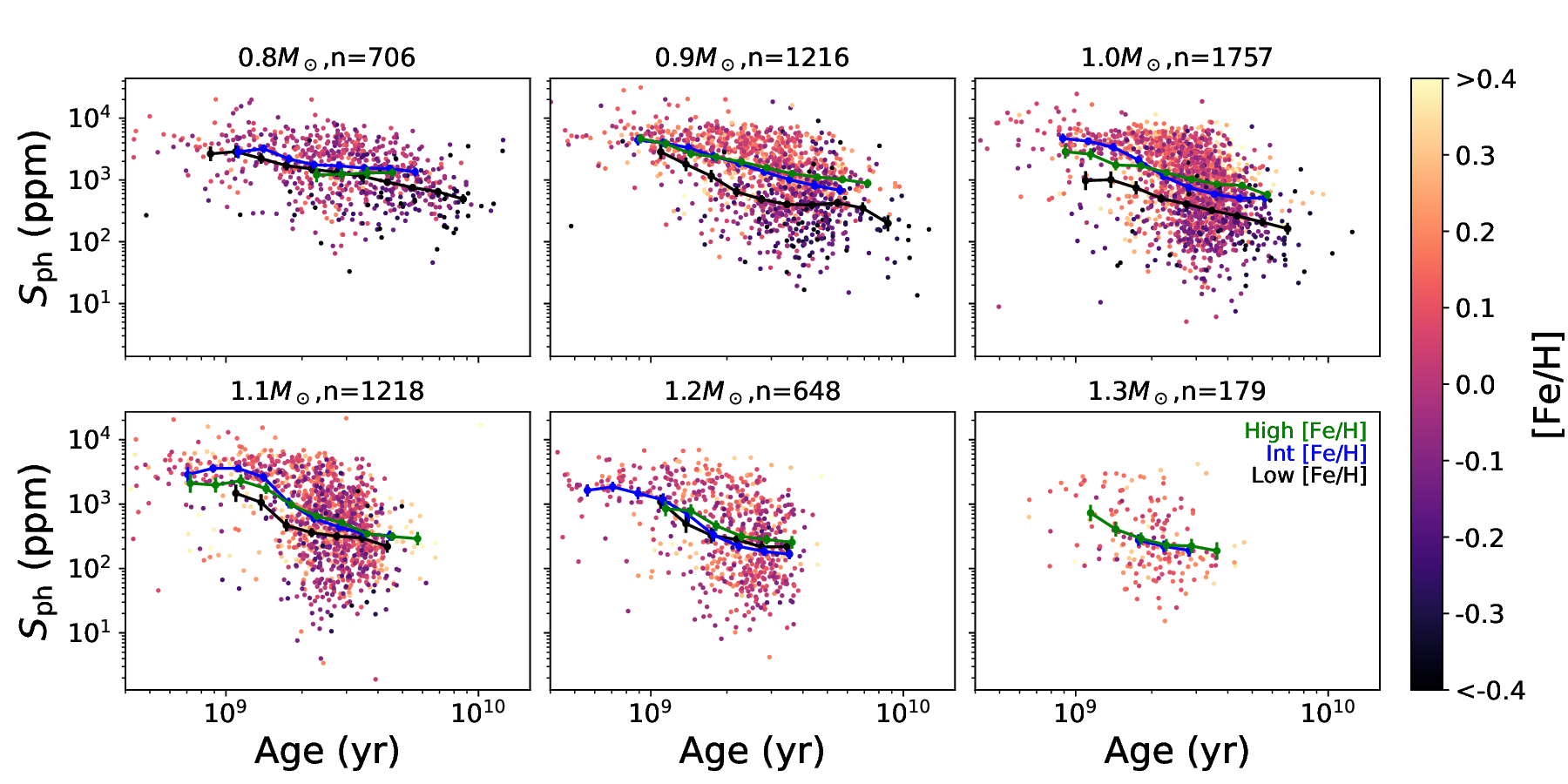}
    \caption{Same as fig. \ref{fig:ActAge} but only for stars that have not yet reached the weakened magnetic braking regime, i.e. those with $\rm Ro \le 1.14$.}
    \label{fig:ActAge_RossCut}
\end{figure*}

In figs. \ref{fig:PeriodAge_RossCut}, \ref{fig:RossbyAge_RossCut} and \ref{fig:ActAge_RossCut}, we show the same plots as figs. \ref{fig:PeriodAge}, \ref{fig:RossbyAge} and \ref{fig:ActAge} respectively but with the stars in the weakened magnetic braking regime removed, i.e. those with $\rm Ro > Ro_{crit}$. Comparing the figs. \ref{fig:PeriodAge_RossCut}, \ref{fig:RossbyAge_RossCut} and \ref{fig:ActAge_RossCut} to figs. \ref{fig:PeriodAge}, \ref{fig:RossbyAge} and \ref{fig:ActAge}, we see that every mass bin has at least some stars in the weakened magnetic braking regime although they are concentrated in the higher mass bins. We also see that the metallicity trends we identified in section \ref{sec:Results} are largely unaffected and, therefore, our conclusions are unaffected as well. Indeed, comparing figs. \ref{fig:PeriodAge_RossCut} and \ref{fig:PeriodAge}, some of the metallicity trends associated with the rotation evolution may actually be clearer once the stars in the weakened magnetic braking regime are removed. Most notably, it was not obvious that there was a metallicity trend in the 1.3$M_\odot$ bin of fig. \ref{fig:PeriodAge} with the running average for the low metallicity stars lying between the the running averages for the intermediate and high metallicity stars. However, low metallicity stars have shorter turnover times and larger Rossby numbers. They are, therefore, more likely to be in the weakened magnetic braking regime. The 1.3$M_\odot$ bin in fig. \ref{fig:PeriodAge_RossCut} shows that once the stars in the weakened magnetic braking regime are removed, the metallicity trend is now clear since the running average for the low metallicity stars is no longer there.



\bibliographystyle{mnras}
\bibliography{MetallicityKinematic} 








\bsp	
\label{lastpage}
\end{document}